\newcommand{\beq}{\begin{equation}}
\newcommand{\eeq}{\end{equation}}
\newcommand{\bqa}{\begin{eqnarray}}
\newcommand{\eqa}{\end{eqnarray}}
\documentclass[12pt]{article}
\usepackage{epsfig}
\def\square{\vcenter{\vbox{\hrule height.4pt
          \hbox{\vrule width.4pt height8pt
          \kern8pt\vrule width.4pt}\hrule height.4pt}}}

\begin{document}

\begin{flushleft}\hspace{11cm}
OHSTPY--HEP--T--97--14\\ \hspace{11cm}
hep--ph/9709418 \\ \hspace{11cm}
\today  \\
\end{flushleft}

\vskip 10mm

\centerline{\Large\bf The 3d Effective Field Theory for}
\centerline{\Large\bf Finite Temperature Scalar Electrodynamics}
\vskip 10mm
\centerline{Jens O. Andersen}
\centerline{\it Department of Physics, The Ohio State University,
Columbus, OH 43210}
\vskip 3mm

\begin{abstract}
{\footnotesize 
The effective field theory approach to high temperature field theory
can be used to study the phase transition in theories with spontaneously
broken symmetry.
I construct a sequence of two effective three--dimensional field theories 
which are valid on successively longer distance scales for a specific model:
model:
$N$ charged scalars coupled to a $U(1)$ gauge field.
The resulting effective Lagrangian can be used to investigate the phase
transition, in particular the order of the phase transition
as a function of $N$, using lattice simulations.}\\ \\ 
PACS number(s): 11.10.Wx, 12.20.DS, 12.38.Bx 
\end{abstract}
\pagebreak
\section{Introduction}
Many problems in quantum field theory at high temperature have been studied
extensively since the work of Dolan and Jackiw on symmetry breaking
almost twenty five years ago~\cite{dolan}.
There has been tremendous progress in perturbative calculations and the methods
available for the investigations of quantum systems at high temperature.

Much of the progress has come from separating the effects of different scales.
The important scales include the temperature $T$ and the scale $gT$
associated with screening 
lengths and quasi--particle masses. Naive perturbation
theory breaks down for soft external momentum $k$ 
($k\sim gT$)
because leading order
results for some physical quantities (e.g. the gluon damping rate)
get contributions from all orders in the loop expansion.
This problem can be solved by integrating out the scale $T$, which leads
to a resummed perturbation theory, which is mainly due to Braaten and 
Pisarski~\cite{pis}. Resummation is a reorganization of the ordinary 
perturbation
expansion in which all effects of the scale $T$ are absorbed into parameters
that appear in effective propagators and effective vertices.
This reorganization is necessary to do consistent perturbative calculations
of real--time processes at high temperature~\cite{pis}.

In the calculation of static quantities such as the 
free energy (or the effective potential) and
screening masses, resummation is just a matter of 
using an effective propagator. 
Resummed perturbation theory has been used extensively as a tool 
for investigating phase transitions at finite temperature.
Hebecker has calculated the two--loop effective potential in the
Abelian Higgs Model~\cite{arthur1} and this model has also been investigated
by  Amelino--Camelia using the composite operator method
(Ref.~\cite{bambino} and Refs. therein).
Fodor and Hebecker have obtained the 
effective potential in the standard model, also in the two--loop
approximation~\cite{fodor}.
In the case of computing static quantities, there exists a simplified
resummation scheme due to Arnold and Espinosa, in which
only the static modes are dressed by thermal masses~\cite{arnold12}.
Both the Abelian Higgs Model and the Standard Model have been subjects of
investigation using this simplified resummation scheme~\cite{arnold12}.
Resummation has also been used to calculate the free energy in 
$g^2\Phi^4$--theory [7-8], QED [9-11] and QCD [11,12]. 

Finally, the electroweak phase transition has
also been studied by lattice simulations directly in four dimensions [13--15].

The strategy of separating scales has proved to be very useful for studying
static properties of high temperature field theories. A very powerful
method for separating scales is effective field theory~\cite{lepage}.
The general idea is to take advantage of two or more well separated mass
scales in the problem and treating one scale at a time.
This is done by constructing a sequence of effective field theories 
which are valid on successively longer distance scales and whose coefficients
encode the short--distance physics.

For hot matter, the nonzero Matsubara modes provide the scale $T$,
while the static modes provide the scale $gT$ and in some cases 
(e.g. in nonabelian gauge theories) the scale $g^2T$ as well~\cite{gross}.
This suggests that one integrates out the nonstatic Matsubara
modes to obtain an effective field theory for the zero modes.
This effective theory is three--dimensional and the process is the
well known dimensional reduction of high temperature field theories [18--20].

The effective field theory approach has been developed into a tool for
quantitative calculations by Farakos, Kajantie, Rummukainen and 
Shaposhnikov~\cite{rummu},
and independently by Braaten and Nieto~\cite{braaten}.
The idea is that one writes down the most general effective three--dimensional
Lagrangian consistent with the symmetries at high temperature.
The coefficients in the effective theory is determined by requiring that
static correlators in the full theory
are reproduced to some desired accuracy by the corresponding correlators
in the effective theory at long distances $R\gg 1/T$.
Moreover, if the effective three--dimensional field theory contains
the two momentum scales $gT$ and $g^2T$, one constructs
a sequence of two effective field theories by matching correlators
at distances $R\gg 1/(gT)$ [21--23].

The effective field theory approach has been used by 
Farakos {\it et al}. [21,24] and by Kajantie {\it et al.} [25--26]
for investigating the important electroweak phase transition which
took place in the early Universe.
One of the reasons for the interest in the electroweak phase transition
is that the baryon asymmetry we observe today, could be a remnant from the
phase transition [27--28].
For electroweak matter at temperatures around $T_c$ there is a hierarchy
of three momentum scales, and so the first step is to construct a sequence
of two effective field theories~\cite{rummu}. 
The second step is the application
of the three--dimensional effective field theory. 
Normally, the perturbation expansion breaks down at temperatures close
to $T_c$, so one must use nonperturbative methods such as lattice simulations.
This has been carried out in [24--28].
The conclusion of their investigation is that the electroweak phase transition
in the Standard Model is not sufficiently strongly first order for realistic
values of the Higgs mass to produce the present baryon asymmetry, and 
one must consider extensions such as the Minimal Supersymmetric Standard
Model and the Two Higgs Doublet Model~[21,24--26,29--32]. 
See also Ref.~\cite{raja}, where these methods have been applied to
$SU(5)$.

Effective field theory methods have been used
by Braaten and Nieto~\cite{braaten2}
to solve the
long--standing
infrared catastrophe of QCD~\cite{linde}.
It is a well--known fact that 
the free energy of 
nonabelian gauge theories cannot
be calculated beyond fifth order in the coupling using resummed 
perturbation theory. 
The method 
breaks down at order $g^{6}$, due to infrared divergences,
as first pointed out by Linde~\cite{linde}. 
These divergences arise from regions where all internal energies
vanish, and so the singularities are the same as in
three--dimensional pure QCD. Thus, the breakdown of perturbation
theory simply reflects the infrared problems 
appearing in a perturbative treatment of any nonabelian
gauge theory in three dimensions.
Using the effective field theory approach 
one can compute order by order in the gauge coupling $g$
the contributions
to the free energy, although some coefficients must be evaluated numerically.
The infrared problems can naturally be avoided if one uses lattice 
simulations directly in four dimensions. However, this is extremely
time consuming in comparison with three--dimensional calculations
and the time savings here
arise from
the reduction of the problem from four to three dimensions, and 
also by integrating out the fermions.

Effective field theory has also been used to organize perturbative
calculations of the free energy in $g^2\Phi^4$--theory~\cite{braaten},
QED~\cite{jens} and SQED~\cite{jenstese}. Moreover, it has also been used
for carrying out perturbative calculations of screening lengths in the
same theories [22,35--36].

Phase transitions, in particular the electroweak phase transition have
been investigated by other methods as well. One of these methods is the
$\epsilon$--expansion. Here, one solves the theory (perturbatively)
in $4-\epsilon$ dimensions and extrapolate the results to $\epsilon =1$
at the end (having assumed there exists a $3d$ description of the system).
The $\epsilon$--expansion combined with renormalization group methods
have been applied in Refs. [37--40]. 

In the present work we apply effective field theory methods to construct
an effective three--dimensional field theory which can be used in the study
of phase transitions for a specific field theory:
a $U(1)$ gauge field coupled to $N$ charged scalars.
This is simply scalar electrodynamics where the scalar field is an 
$N$--component complex vector, and we shall refer to this model as 
SQED in the following. For $N=1$, the effective field theory has
already been constructed 
in 
Ref.~\cite{rummu}. Several aspects of the phase transition
have been studied numerically in Refs. [41-51].
Moreover, this 
theory has previously been investigated by Arnold~\cite{epsilon} and
Lawrie~\cite{law} using the $\epsilon$--expansion. 
The renormalization group equations have a nontrivial
infrared fixed point in $4-\epsilon$ dimensions for $N>N_c$, where 
$N_c\approx 365.9$ [37--38].
Such fixed points are taken as evidence for a second order phase transition,
since the theory looks the same on all distance scales~\cite{arnold12}
and so this suggests that the order of the phase transition depends on $N$.
According to Ref.~\cite{epsilon}, the $\epsilon$--expansion is not so well
behaved when the number of fields $N$ becomes large. So it is of 
interest to study this system by other means. The results presented in this
work provide a first step in this direction.

The plan of the article is as follows.
In section II
we review the ideas behind dimensional reduction and the construction of 
effective three--dimensional field theories.
In section III and IV we determine the coefficients 
in the two effective field theories arising in our model.
In section V we summarize and conclude.
In Appendix A and B, the notation  and conventions are given.
We also list the sum--integrals in the full theory as well as the integrals
in the effective theories which are needed in the present work.
\section{Dimensional Reduction and Effective Field Theories}
In this section we briefly discuss the ideas behind dimensional reduction and 
the effective field theory approach to phase transitions at finite 
temperature. 

The specific model we study in the present work consists of $N$
complex scalar fields coupled to an Abelian gauge field.
The Euclidean Lagrangian for SQED reads
\beq
\label{sqed}
{\cal L}_{\mbox{\scriptsize SQED}}
=\frac{1}{4}F_{\mu\nu}F_{\mu\nu}+({\cal D_{\mu}}\Phi )
^{\dagger}({\cal D_{\mu}}\Phi )+\nu^{2}\Phi^{\dagger}\Phi
+\frac{\lambda}{6} (\Phi^{\dagger}\Phi)^{2}+
{\cal L}_{\mbox{\footnotesize gf}}+
{\cal L}_{\mbox{\footnotesize gh}}.
\eeq
Here $D_{\mu}=\partial_{\mu}+ieA_{\mu}$ is the covariant derivative
and
$\Phi^{\dagger}=(\Phi_1^{\dagger},\Phi_2^{\dagger},...,\Phi_N^{\dagger})$.
$\Phi$ is the corresponding column vector.
In the present work, we perform the calculations in the Landau gauge. 
This is merely a convenient choice, since many of the diagrams
vanish in this gauge. Our final results are of course gauge fixing
independent.

In the imaginary time formalism bosonic fields are
periodic in the time direction with period $\beta$, while fermionic fields
are antiperiodic with the same period.  This allows one to decompose
the fields into their Fourier components, which are characterized by their
Matsubara frequencies. For bosonic fields these are $2n\pi T$ and
for fermions they are $(2n+1)\pi T$.
The Matsubara frequencies act as masses for the Fourier components of the
fields, and one can view a four--dimensional field theory at finite
temperature as a three--dimensional field theory at zero temperature
with an infinite tower of fields~\cite{lands}.
Thus, the nonzero Matsubara modes have masses of order $T$, while the 
static mode of $A_0$
acquires a thermal mass of order $gT$.
The zero mode of $A_i$
is massless (there is no magnetic mass in Abelian gauge theories)
and if the temperature is close to the critical temperature the 
static modes of the scalar
fields have masses of order $g^2T$.
Hence, there is a hierarchy of three momentum scales, $T$, $gT$ and $g^2T$
which are well separated in the weak coupling limit.
This suggests that one construct a sequence of two effective field theories
which are valid on successively longer distance scales:
The first step is to integrate out the nonzero Matsubara
frequencies and construct an effective three--dimensional field theory
for the $n=0$ bosonic modes. This is the familiar process of 
dimensional reduction of hot field theories [18--22].
The second effective field theory is obtained
by integrating out the timelike component of the gauge 
field [21--23]. 

The first effective field theory is called electrostatic scalar electrodynamics
(ESQED) and the fields can be approximately identified with the zero--frequency
modes of the original fields.
${\cal L}_{\mbox{\scriptsize ESQED}}$ consists of a real massive
scalar field, which can be identified with the zero mode of the temporal
component of the gauge field. We denote this field by $\rho$.
Moreover, we have the $N$--component scalar field 
$\phi$ and the three--dimensional gauge field
$A_i^{3d}$ which are associated with the zero--frequency modes of 
$\Phi$ and $A_i$ in SQED, respectively. We can then schematically write
\beq
\label{rel}
\phi ({\bf x})\approx\sqrt{T}
\int_0^{\beta}d\tau\Phi({\bf x},\tau)\,,\hspace{0.4cm}
A_i^{3d}({\bf x})\approx\sqrt{T}\int_0^{\beta}
d\tau A_i({\bf x},\tau)\,,\hspace{0.4cm}
\rho ({\bf x})\approx\sqrt{T}\int_0^{\beta}d\tau A_0({\bf x},\tau)\,.
\eeq
The symmetries are as follows: There is a gauged $U(1)$ symmetry of
$\phi$ and a $Z_2$--symmetry of $\rho$.
Hence, the Lagrangian of ESQED is
\bqa\nonumber
\label{esqed}
{\cal L}_{\mbox{\scriptsize ESQED}}
&=&\frac{1}{4}F_{ij}F_{ij}+({\cal D}_{i}\phi )
^{\dagger}({\cal D}_{i}\phi )+M^{2}(\Lambda)\phi^{\dagger}\phi
+\frac{\lambda_E(\Lambda)}{6} (\phi^{\dagger}\phi)^{2} 
+\frac{1}{2}(\partial_i\rho)^2\\
&&+\frac{1}{2}m_{E}^{2}(\Lambda)\rho^2
+\frac{\lambda_{A}(\Lambda)}{24}\rho^4
+h^{2}_E(\Lambda)\phi^{\dagger}\phi\rho^{2}
+{\cal L}_{\footnotesize \mbox{gf}}
+{\cal L}_{\mbox{\footnotesize gh}}+\delta{\cal L}.
\eqa
The parameters in ESQED are called {\it short--distance coefficients}.
The term $\delta {\cal L}_{\mbox{\scriptsize ESQED}}$ represents all other
terms in ESQED which can be constructed out of the fields and which respect
the symmetries. Examples of such terms are $h(\Lambda)\rho^2F_{ij}^2$
and $g(\Lambda)(\phi^{\dagger}\phi)^{3}\rho^2$.

The second three--dimensional
effective field theory is named magnetostatic scalar
electrodynamics (MSQED) and consists of the fields $\tilde{\phi}$ 
and $\tilde{A}_i^{3d}$. The fields in MSQED are to a first approximation
identified with the fields in ESQED.
The symmetry is a gauged $U(1)$ symmetry, exactly as in full SQED.
The Lagrangian of MSQED then reads
\beq
\label{lmsqed}
{\cal L}_{\mbox{\scriptsize MSQED}}=
\frac{1}{4}F_{ij}F_{ij}+({\cal D}_{i}\tilde{\phi} )
^{\dagger}({\cal D}_{i}\tilde{\phi} )+\tilde{M}^{2}(\Lambda)
\tilde{\phi}^{\dagger}\tilde{\phi}
+\frac{\lambda_M(\Lambda)}{6} (\tilde{\phi}^{\dagger}\tilde{\phi})^{2}
+{\cal L}_{\mbox{\footnotesize gf}}
+{\cal L}_{\mbox{\footnotesize gh}}
+\delta{\cal L}_{\mbox{\scriptsize MSQED}}.
\eeq
The parameters of MSQED are termed {\it middle--distance coefficients}.
The term $\delta {\cal L}_{\mbox{\scriptsize MSQED}}$ includes all 
operators that can be made out of $\tilde{A}_i$ and $\tilde{\phi}$, 
for instance
$c(\Lambda)(F_{ij}F_{ij})^2$.

In the equations above, we have indicated that the parameters generally
depend on $\Lambda$, which is the ultraviolet cutoff of the effective
theory. This cutoff dependence is necessary in order to cancel the 
$\Lambda$-dependence which arises in perturbative calculations using the 
effective theory.

Matching static Greens functions in SQED and ESQED is complicated
by the breakdown of the relation (\ref{rel}) between the fields in the
fundamental theory and the fields in the effective theory. 
In the present case this breakdown takes place at leading
order in $e^2$ and we must allow 
for short--distance coefficients multiplying
the fields $\phi$, $A_i^{3d}$ and $\rho$ in ESQED~\cite{braaten}.
These short--distance coefficients are 
associated with field strength renormalization of the fundamental 
fields. They can be found by computing the momentum dependent part of
the propagator of the relevant fields [21,25].
In the one--loop approximation, we denote 
these coefficients by $\Sigma^{(1)\prime}(0)$, 
$\Pi^{(1)\prime}(0)$ and $\Pi_{00}^{(1)\prime}(0)$ 
(see subsection~\ref{feltet}). The
relations between the fields in SQED and ESQED at leading
order in $\lambda$ and $e^2$ are
\bqa
\label{relmod}
\left[1-\Sigma^{(1)\prime}(0)\right]\phi ({\bf x})&\approx&\sqrt{T}
\int_0^{\beta}d\tau\Phi({\bf x},\tau)\,,\\
\left[1-\Pi^{(1)\prime}(0)\right]A_i^{3d}({\bf x})&\approx&\sqrt{T}\int_0^{\beta}
d\tau A_i({\bf x},\tau)\,,\\
\left[1-\Pi_{00}^{(1)\prime}(0)
\right]\rho ({\bf x})&\approx&\sqrt{T}\int_0^{\beta}d\tau A_0({\bf x},\tau)\,.
\eqa
The above remarks also apply when we match correlators in ESQED and MSQED, 
although the middle--distance coefficients vanish at one--loop
(see subsection~\ref{felt2}).
\section{Short--distance Coefficients}
In this section we determine the short--distance coefficients $m^{2}_E(\Lambda)$
and $M^{2}(\Lambda)$ to next--to--leading order in the parameters 
$\nu^2$, $\lambda$ and $e^2$.
We also compute the parameters $\lambda_E(\Lambda)$, $e^2_E(\Lambda)$
and $h^2_E(\Lambda)$ to next--to--leading order,
as well as the 
coefficient $\lambda_A(\Lambda)$ to leading order.

In the present work we shall use naive or 
{\it strict perturbation} theory~\cite{braaten} to determine the parameters
in the effective theory.
The Lagrangian of SQED is split according to 
\bqa \nonumber
\label{strict1}
({\cal L}_{\mbox{\scriptsize SQED}})_{0}&=&\frac{1}{4}
F_{\mu\nu}F_{\mu\nu}+(\partial_{\mu}
\Phi )^{\dagger}(\partial_{\mu}\Phi )
+{\cal L}_{\mbox{\footnotesize gf}}
+{\cal L}_{\mbox{\footnotesize gh}},
\\ 
({\cal L}_{\mbox{\scriptsize SQED}})
_{\mbox{\footnotesize int}}&=&\nu^2\Phi^{\dagger}\Phi+
e^{2}\Phi^{\dagger}\Phi A_{\mu}^{2}
-ieA_{\mu}
(\Phi^{\dagger}\partial_{\mu}\Phi-\Phi\partial_{\mu}
\Phi^{\dagger})
+\frac{\lambda}{6}(\Phi^{\dagger}\Phi)^2.
\eqa
Although the strict perturbation expansion breaks down at distance scales
$R\gg 1/T$, we can use it as device determining the short--distance
coefficients in the effective Lagrangian. The idea is that physical
quantities receive contributions from three momentum scales $T$, $eT$
and $e^2T$. The parameters of ESQED are insensitive to the scales $eT$
and $e^2T$ but 
encode the physics at the scale
$T$.
However, in the matching calculations we must make the same incorrect 
assumptions about the long--distance behaviour in the effective theory.
If we tune the parameters so that the two theories are equal at
long distances, then the infrared divergences in full SQED are
identical to those encountered in ESQED. 
Of course, in perturbative calculations, one must
regularize the infrared divergences by an infrared cutoff.
In the present work dimensional regularization is used. 
In the effective
theory these incorrect assumptions amount to treating the mass parameters
as well as other operators as perturbations. 
Strict perturbation theory is then
defined by the following decomposition of
the Lagrangian of ESQED
\bqa\nonumber
({\cal L}_{\mbox{\scriptsize ESQED}})_0
&=&\frac{1}{4}F_{ij}F_{ij}+(\partial_{i}
\phi^{\dagger})(\partial_{i}\phi )
+\frac{1}{2}(\partial_{i}\rho)^{2}+
{\cal L}_{\mbox{\footnotesize gf}}+
{\cal L}_{\mbox{\footnotesize gh}},
\\ \nonumber
({\cal L}_{\mbox{\scriptsize ESQED}})_{\mbox{\footnotesize int}}
&=&M^{2}(\Lambda )\phi^{\dagger}
\phi+\frac{1}{2}
m^{2}_E(\Lambda )\rho^{2}+
\frac{\lambda_E(\Lambda)}{6} (\phi^{\dagger}\phi)^{2} +
e^{2}_{E}(\Lambda)\phi^{\dagger}\phi A_{i}^{2}+\\
&&
h_E^2(\Lambda)\phi^{\dagger}\phi\rho^{2}+ie_{E}(\Lambda)A_{i}
(\phi^{\dagger}\partial_{i}\phi-\phi\partial_{i}
\phi^{\dagger})+\frac{\lambda_A(\Lambda)}{24}\rho^4+\delta{\cal L}.
\eqa
In full SQED, wiggly and solid lines denote the propagators
of photons and charged scalars, respectively.
In ESQED, the same conventions apply. Moreover,
dashed lines denote the propagators
of the real scalar field $\rho$.
A cross in the Feynman diagrams denotes the insertion of the operator $\nu^2$. 
Note also that the figures only display those diagrams in the perturbative
expansion which are 
non--vanishing in the Landau gauge.
\subsection{Field Normalization Constants}\label{feltet}  
In this subsection we compute the short--distance coefficients 
$\Sigma^{(1)\prime}(0)$, 
$\Pi^{(1)\prime}(0)$ and $\Pi_{00}^{(1)\prime}(0)$
which
multiply the fields $\phi$, $A_{i}^{3d}$ and $\rho$ in ESQED.

We denote the {\it static} self--energy function of the scalar field by
$\Sigma ({\bf k})$, and the static polarization tensor of the gauge field
by $\Pi_{\mu\nu}({\bf k})$. Now, 
$\Sigma ({\bf k})$ and $\Pi_{\mu\nu}({\bf k})$
can be expanded in number of loops in the loop expansion
and can also be expanded in powers of the external momentum ${\bf k}$.
If we denote the nth order contribution to the scalar self--energy function
by $\Sigma^{(n)} ({\bf k})$, we can write
\beq
\Sigma ({\bf k})=\Sigma^{(1)}(0)+k^2\Sigma^{(1)\prime}(0)+
\Sigma^{(2)}(0)+....
\eeq
Here, and in the rest of the paper $k=|{\bf k}|$.
The one--loop diagrams contributing to the self--energy of the scalar field
is shown in Fig.~\ref{gor}, and read
\bqa\nonumber
\Sigma^{(1)} ({\bf k})&=&\frac{(N+1)\lambda\nu^2}{3}
\hbox{$\sum$}\!\!\!\!\!\!\int_P\frac{1}{P^4}+
\frac{1}{3}\left[(N+1)\lambda
+3(d-1)e^2\right]\hbox{$\sum$}\!\!\!\!\!\!\int_P\frac{1}{P^2}\\
&&
-4e^2\hbox{$\sum$}\!\!\!\!\!\!\int_P
\frac{k^2}{P^2(P+K)^2}
+4e^2\hbox{$\sum$}\!\!\!\!\!\!\int_P\frac{({\bf p}{\bf k})^2}{P^4(P+K)^2}.
\eqa
Expanding in powers of the external momentum ${\bf k}$ gives
\bqa
\label{avles}
\Sigma^{(1)} ({\bf k})&=&
\frac{1}{3}\left[(N+1)\lambda\nu^2-9e^2k^2\right]
\hbox{$\sum$}\!\!\!\!\!\!\int_P\frac{1}{P^4}
+\frac{1}{3}\left[(N+1)\lambda
+3(d-1)e^2\right]\hbox{$\sum$}\!\!\!\!\!\!\int_P\frac{1}{P^2}
+....
\eqa
From~(\ref{avles}) we immediately get the unrenormalized coefficients
\bqa
\label{sjoel1}
\Sigma^{(1)} (0)&=&
\frac{\nu^2(N+1)\lambda}{3}
\hbox{$\sum$}\!\!\!\!\!\!\int_P\frac{1}{P^4}+
\frac{1}{3}\left[(N+1)\lambda
+3(d-1)e^2\right]\hbox{$\sum$}\!\!\!\!\!\!\int_P\frac{1}{P^2},\\
\label{sjoeld}
\Sigma^{(1)\prime} (0)&=&
-3e^2\hbox{$\sum$}\!\!\!\!\!\!\int_P\frac{1}{P^4}.
\eqa
The sum--integral in~(\ref{sjoeld}) is ultraviolet divergent and the divergence
is removed by the field strength renormalization counterterm, To leading
order we have
\beq
Z_{\Phi}=1+\frac{3e^{2}}{16\pi^2\epsilon}.
\eeq
We then obtain 
\bqa
\label{gdep}
\Sigma^{(1)\prime}(0)&=&\frac{3e^{2}}{16\pi^{2}}\left[\ln\frac{\Lambda}{4\pi T}+\gamma_E\right].
\eqa
Let us next move to the gauge field. The one--loop diagrams which contribute
to the photon polarization tensor are displayed in 
Fig.~\ref{aure}. The calculations are straightforward, and one finds
%
\bqa
\Pi_{00}^{(1)\prime}(0)&=&
\frac{Ne^{2}}{48\pi^{2}}\left[\ln\frac{\Lambda}{4\pi T}+\gamma_E+1\right],\\
\Pi^{(1)\prime}(0)&=&
\frac{Ne^{2}}{48\pi^{2}}\left[\ln\frac{\Lambda}{4\pi T}+\gamma_E\right].
\eqa
\subsection{Coupling Constants}\label{previ}
In this subsection we present the results for the coupling constants
$\lambda_E(\Lambda)$, $e^2_E(\Lambda)$ and $h^2_E(\Lambda)$
to next--to--leading order in the coupling constants of the full theory.
We also give the result for
the coupling constant $\lambda_A(\Lambda)$ in the
one--loop approximation.
For $N=1$ these parameters have been calculated in~\cite{rummu}.

Let us first consider the coefficient $\lambda_E(\Lambda)$. 
To leading order one can simply read off this parameter from
the full theory. Substituting $\Phi({\bf x},\tau)\rightarrow
\sqrt{T}\phi({\bf x})$
into~(\ref{sqed})
and comparing $\int_0^{\beta}d\tau{\cal L}_{\mbox{\scriptsize SQED}}$ 
with the Lagrangian of ESQED we find
\beq
\lambda_E(\Lambda)=\lambda T.
\eeq
One way to calculate the coupling $\lambda_E(\Lambda)$ beyond leading order,
is by matching the static four--point function of the Higgs field 
in full SQED with the four--point function of the Higgs field in ESQED.
This is complicated by the breakdown of the relation~(\ref{rel}).
At next--to--leading order it is sufficient to take into account the
short--distance coefficient which multiplies $\phi$.
 
\begin{figure}[b]
\underline{FIGURE CAPTIONS:}
\caption{One--loop scalar self-energy diagrams in the full theory.}
\label{gor}
\caption{One--loop diagrams for the photon polarization tensor in the full theory.}
\label{aure}
\caption{One-loop graphs contributing to the scalar four-point function
in the full theory.}
\label{gjedde}
\caption{\protect One--loop diagrams needed for the calculating the couplings
$e^2_E(\Lambda)$ and $h^2_E(\Lambda)$.}
\label{fugl}
\caption{\protect One-loop diagrams contributing to the four--point
function of $\rho$ in SQED.}
\label{bever}
\caption{Two-loop scalar self--energy diagrams in SQED.}
\label{ape}
\caption{Two-loop self--energy diagrams for the timelike component
of the gauge field in SQED.}
\label{bille}
\caption{\protect One and two--loops diagram contributing to the
scalar self--energy function in ESQED.}
\label{hjort}
\caption{One--loop diagram relevant for the calculation of scalar 
self--coupling in ESQED.}
\label{tryte}
\end{figure}
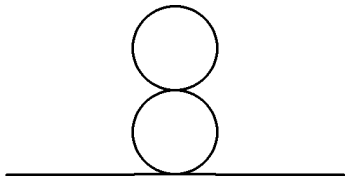
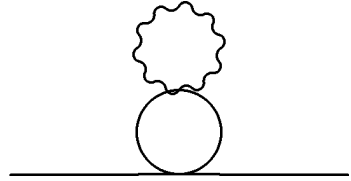
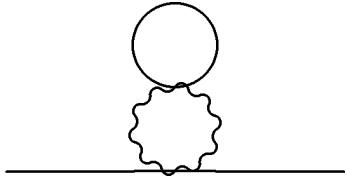
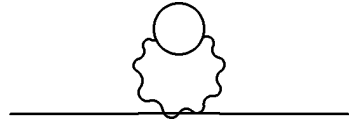
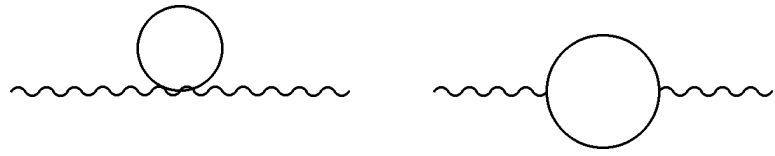
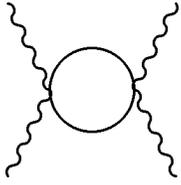
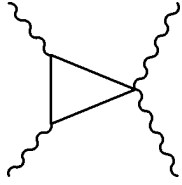
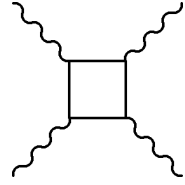
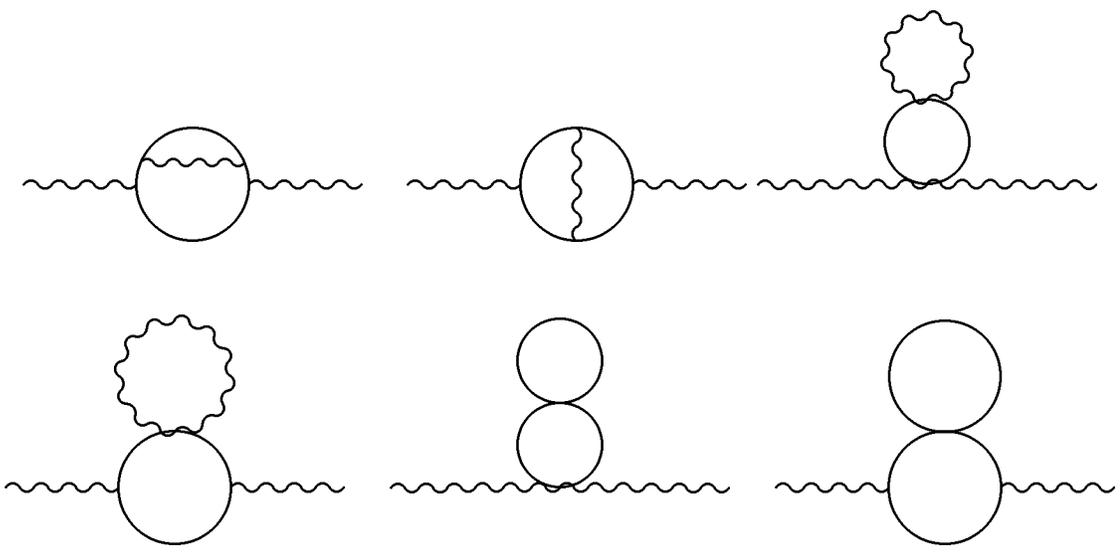
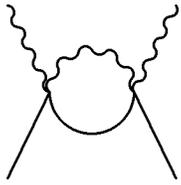
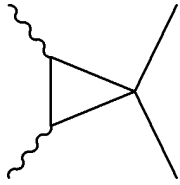
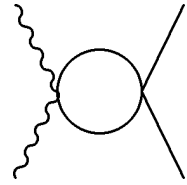
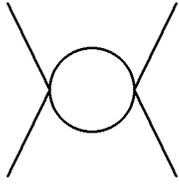
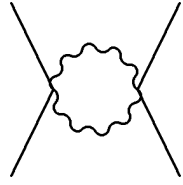
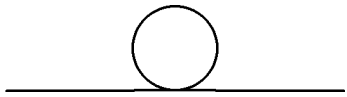
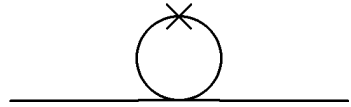
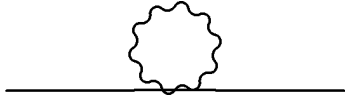
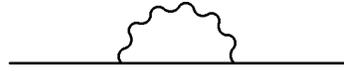
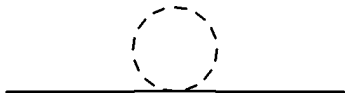
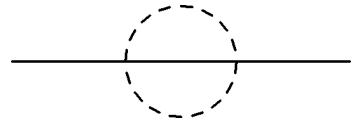
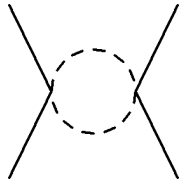

We denote the four--point of the Higgs field in SQED by 
$\Gamma^{\mbox{\scriptsize SQED}}_{\phi_1,\phi_1,\phi_1,\phi_1}({\bf k})$, where ${\bf k}$
collectively denotes the external momenta. The one--loop correction to the
four--point function is given by the Feynman diagrams in Fig.~\ref{gjedde}.
Taken
at zero external momenta, one finds
\beq
\Gamma^{\mbox{\scriptsize SQED}}_{\phi_1\phi_1\phi_1\phi_1}(0)=\lambda
-\frac{1}{3}\left[(N+4)
\lambda^2
+18(d-1)e^4
\right]\hbox{$\sum$}\!\!\!\!\!\!\int_{P}\frac{1}{P^{4}}.
\eeq
In ESQED, we denote the corresponding four--point function by
$\Gamma^{\mbox{\scriptsize ESQED}}_{\phi_1,\phi_1,\phi_1,\phi_1}({\bf k})$.
Since all the fields are massless in the strict perturbation
expansion and all diagrams are taken at vanishing external momenta, there is
no scale in the integrals. Thus the loop corrections to 
$\Gamma^{\mbox{\scriptsize ESQED}}_{\phi_1,\phi_1,\phi_1,\phi_1}(0)$ vanish:
\beq
\Gamma^{\mbox{\scriptsize ESQED}}_{\phi_1,\phi_1,\phi_1,\phi_1}(0)=\lambda_E(\Lambda). 
\eeq
Taking into account the short--distance coefficient multiplying the field
$\phi$, the matching leads to the following equation
\beq
\label{renormierung}
\lambda_E(\Lambda)=\lambda T
-\frac{1}{3}\left[(N+4)\lambda^2
-18\lambda e^2
+18\left(d-1\right)e^4\right]T\hbox{$\sum$}\!\!\!\!\!\!\int_{P}\frac{1}{P^{4}}.
\eeq
Renormalization of the quartic coupling $\lambda$ 
is carried out by the substitution
$\lambda\rightarrow Z_{\lambda}\lambda$ in the first term on the right hand
side of~(\ref{renormierung}), where
\beq
\label{lambdi}
Z_{\lambda}=1+\frac{(N+4)\lambda -18\lambda e^2+54e^4}{48\pi^{2}\epsilon}.
\eeq
This yields 
\bqa
\label{lambdae}
\lambda_E(\Lambda)=T\left[\lambda -
\frac{(N+4)\lambda^2-18\lambda e^2+54e^4}{24\pi^2}\left(
\ln\frac{\Lambda}{4\pi T}+\gamma_E\right)+\frac{3e^4}{4\pi^2}\right].
\eqa
The couplings $e^{2}_{E}(\Lambda)$ 
and $h_{E}^2(\Lambda)$ are computed by 
matching the correlators
$\Gamma^{\mbox{\scriptsize SQED}}_{\Phi_1^{\dagger}\Phi_1 A_iA_j}({\bf k})$ and
$\Gamma^{\mbox{\scriptsize SQED}}_{\Phi_1^{\dagger}\Phi_1 A_0A_0}({\bf k})$ in full SQED
with the corresponding correlators in ESQED.
The relevant diagrams are displayed in Fig.~\ref{fugl}
and the results are:
\bqa
\label{ekopling}
e_{E}^{2}(\Lambda)&=&e^{2}T\left[1-\frac{Ne^2}{24\pi^2}\left(
\ln\frac{\Lambda}{4\pi T}+\gamma_E\right)\right],\\
h_E^2\left(\Lambda\right)&=&
e^{2}T\left[1-\frac{Ne^2}{24\pi^2}\left(
\ln\frac{\Lambda}{4\pi T}+\gamma_E+1\right)+\frac{(N+3)\lambda}{48\pi^2}+
\frac{e^2}{8\pi^2}\right].
\eqa
We close this subsection by giving the coefficient in front
of the operator $\rho^{4}$. 
To leading order in the couplings of full SQED, $\lambda_A(\Lambda)$
is given by the one--loop contribution to the
four--point function for timelike photons at zero external
momenta. This correlator is denoted by
$\Gamma^{\mbox{\scriptsize SQED}}_{A_{0}A_{0}A_{0}A_{0}}({\bf k})$. 
The one--loop graphs contributing to this correlator are displayed
in Fig.~\ref{bever} and one finds: 
\beq
\lambda_{A}(\Lambda)=\frac{Ne^{4}T}{\pi^{2}}.
\eeq
The four coupling constants 
$\lambda_E(\Lambda)$, $e^{2}_{E}(\Lambda)$, $h_E^{2}(\Lambda)$ 
and $\lambda_{A}(\Lambda)$ are independent of the cutoff $\Lambda$
at next--to--leading order in the coupling constants of SQED. This follows
directly from the RG--equations for $\lambda$ and $e^2$:
\bqa
\label{renorm1}
\mu\frac{d\lambda}{d\mu}&=&
\frac{(N+4)\lambda^2-18\lambda e^2+54e^4}{24\pi^2},\\
\label{renorm2}
\mu\frac{de^2}{d\mu}&=&\frac{Ne^4}{24\pi^2},
\eqa
Thus, we can trade the scale $\Lambda$ for an arbitrary renormalization 
scale $\mu$. 
\subsection{Mass Parameters}
In this subsection we calculate the mass parameters $M^2(\Lambda)$
and $m^2_E(\Lambda)$ 
in the effective Lagrangian
at next--to--leading order in $\nu^2$, $\lambda$ and $e^2$. 
The leading order results for $N=1$ can be found in e.g.~\cite{rummu}, while
the result for $m^2_E(\Lambda)$ at next--to--leading order has been obtained
in~\cite{jenstese}.
There are several ways of determining the mass parameters. One way is
to match the propagator of the zero--frequency mode in the full theory with
the propagator in ESQED.

Let us denote the static two--point function
of the Higgs field in SQED 
by $\Gamma^{\mbox{\scriptsize SQED}}_{\phi_1\phi_1}({\bf k})$. 
Since we can expand the self--energy function
in both powers of the external momentum and number of loops we can write
\beq
\Gamma^{\mbox{\scriptsize SQED}}_{\phi_1\phi_1}({\bf k})
=k^2+\nu^2+\Sigma^{(1)}({\bf k})
+k^2\Sigma^{(1)\prime}({\bf k})
+\Sigma^{(2)}({\bf k}).
\eeq
Similarly, we denote the two--point function of the Higgs field in ESQED
by $\Gamma^{\mbox{\scriptsize ESQED}}_{\phi_1\phi_1}({\bf k})$. We can then write
\beq
\Gamma^{\mbox{\scriptsize ESQED}}_{\phi_1\phi_1}({\bf k},\Lambda)=k^2+M^2(\Lambda)
+\delta M^2.
\eeq 
Here, we have added a mass counterterm $\delta M^2$, which is 
associated with mass renormalization.
The matching requirement is then 
\beq
\Gamma^{\mbox{\scriptsize SQED}}_{\phi_1\phi_1}({\bf k})=\left[1+\Sigma^{(1)\prime}(0)\right]
\Gamma^{\mbox{\scriptsize ESQED}}_{\phi_1\phi_1}({\bf k}).
\eeq
The factor $\left[1+\Sigma^{(1)\prime}(0)\right]$ is a consequence
of the short--distance coefficient that multiplies the field
$\phi$.
Solving for the mass parameter, we obtain
\beq
M^2(\Lambda)=-\nu^2\left[1-\Sigma^{(1)\prime}(0)\right]+
\Sigma^{(1)}(0)\left[1-\Sigma^{(1)\prime}(0)\right]
+\Sigma^{(2)}(0)
-\delta M^2.
\eeq
$\Sigma^{(1)}(0)$ and $\Sigma^{(1)\prime}(0)$ are given by~(\ref{sjoel1})
and~(\ref{sjoeld}).
The two--loop contributions to the scalar self--energy are depicted in 
Fig.~\ref{ape} and
yield
\bqa
\Sigma^{(2)}(0)=
-\frac{1}{9}\left[(N+1)^2\lambda^2-3(d-1)(N+1)\lambda e^2+18(d-2)Ne^4
\right]\hbox{$\sum$}\!\!\!\!\!\!\int_{PQ}\frac{1}{P^{2}Q^{4}}.
\eqa
The parameters $\nu^2$ and $e^2$ 
are renormalized by the 
substitutions~\cite{arnold12}
\bqa
Z_{\nu^2}&=&1+\frac{(N+1)\lambda}{48\pi^2\epsilon}
-\frac{3e^2}{16\pi^2\epsilon},\\
\label{charge}
Z_{e^2}&=&1+\frac{Ne^{2}}{48\pi^{2}\epsilon},
\eqa 
while the renormalization constantfor $\lambda$ 
is given by~(\ref{lambdi}). 
We are still left with a pole in $\epsilon$. This divergence is canceled
by the mass counterterm, which thereby is determined to be
\beq
\delta M^{2}=\frac{(N+1)\lambda^{2}T^{2}-6(N+1)\lambda e^2T^2+9(N+5)e^4T^2}
{576\pi^{2}\epsilon}.
\eeq
It is also convenient to express the mass parameter in terms of the
renormalization group invariant
coupling constants of ESQED that we obtained in the previous subsection.
This gives the mass parameter $M^2(\Lambda)$ to two--loop order:
\bqa\nonumber
\label{emass}
M^{2}(\Lambda)&=&\tilde{\nu}^{2}(\mu)+
\frac{1}{36}\left[(N+1)\lambda_E+
9e^2_E\right]T
+\frac{1}{288\pi^2}\left[(N+1)\lambda e^2T^2-
(N+15)e^4T^2\right]
\\&&
-\frac{1}{144\pi^2}
\left[
(N+1)\lambda^2T^{2}-6(N+1)\lambda e^2T^2+9(N+5)e^4
\right]\left[\ln\frac{3T}{\Lambda}+c\right].
\eqa
Here, the renormalization group invariant mass parameter
$\tilde{\nu}^{2}(\Lambda)$ is~\cite{epsilon}
\beq
\tilde{\nu}^{2}(\Lambda)
=\nu^2\Bigg\{1+\frac{1}{48\pi^2}
\left[18e^{2}-2(N+1)\lambda\right]
\left[\ln\frac{\Lambda}{4\pi T}+\gamma_E\right]\Bigg\}
\eeq
and the constant $c$ introduced in~\cite{rummu} is
\bqa\nonumber
c&=&\frac{1}{2}\left[\ln\frac{8\pi}{9}+\frac{\zeta^{\prime}(2)}{\zeta (2)}-2\gamma_E\right]\\
&\approx& -0.348725.
\eqa
The parameters $\lambda_E$ and $e_E^2$ are evaluated at some scale $\mu$
and the remaining dependence on $\Lambda$ shows that
$M^2(\Lambda)$ depends explicitly upon the factorization scale 
$\Lambda$. This is necessary in order to cancel the 
$\Lambda$--dependence which arises in the effective theory.

Let us now turn to the mass parameter $m_E^2(\Lambda)$. 
This parameter is determined by matching the 
propagator of the zero--frequency mode of the timelike component
of the gauge field in SQED with the propagator of the real scalar field $\rho$
in ESQED. 
The one--loop graphs are depicted in Fig~\ref{aure} and
the two--loop part of the
self--energy of $A_0$ is given by the displayed graphs in Fig.~\ref{bille}. 
In complete analogy with the calculations
of the scalar mass parameter, we find 
\beq
m_E^2(\Lambda)=\frac{Ne^{2}T^{2}}{3}\left[1-\frac{2Ne^2}{3(4\pi)^2}\left(
\ln \frac{\Lambda }{4\pi T}+\gamma_{E}+1\right)
+\frac{3e^2}{(4\pi)^2}\right]
+\frac{N(N+1)\lambda e^2T^2}{144\pi^2}.
\eeq
In contrast with the scalar mass parameter, $m_E^2(\Lambda)$ has
no explicit dependence on $\Lambda$. This is easily verified by
using the renormalization group equation~(\ref{renorm2})
for the gauge coupling 
$e^2$.
\section{Middle--distance Coefficients}
In this section we determine the middle--distance coefficients
of MSQED, which is given
by~(\ref{lmsqed}).
We know from general renormalization theory that MSQED can reproduce
the correlators of ESQED at long distances $R\gg 1/eT$
to any desired accuracy
by adding sufficiently many operators and tuning their coefficients 
as functions of the parameters of ESQED.
The middle--distance 
coefficients are sensitive to momentum scales $T$ and $eT$.
The scale $T$ has already been 
encoded in the parameters by the matching which was carried out in the 
previous section. In order to treat the physics on the scale $eT$ correctly, 
we must include the mass parameter $m_E^2(\Lambda)$ in the free 
part of the Lagrangian. By doing this, we treat the effects of $m_E^2(\Lambda$)
to all orders, while the other parameters in ESQED are treated as
perturbations. 
In particular this means that the scalar mass parameter is treated as a
perturbation.
Of course, this way of doing perturbative calculations
is also afflicted with infrared divergences. However, these divergences
are screened at the scale $e^2T$, to which the parameters of MSQED are
insensitive. As long as we make the same incorrect assumptions
about the long--distance behaviour in MSQED, we can use this method to 
determine the middle--distance coefficients of MSQED.

According to the discussion above, the Lagrangian of ESQED is
split into a free and an interacting piece:
\bqa\nonumber
({\cal L}_{\mbox{\scriptsize ESQED}})_{0}
&=&\frac{1}{4}F_{ij}F_{ij}+(\partial_{i}
\phi^{\dagger})(\partial_{i}\phi )
+\frac{1}{2}(\partial_{i}\rho)^{2}
+\frac{1}{2}m^{2}_E(\Lambda )\rho^{2}+
{\cal L}_{\mbox{\footnotesize gf}}+
{\cal L}_{\mbox{\footnotesize gh}},
\\ \nonumber
({\cal L}_{\mbox{\scriptsize ESQED}})_{\mbox{\scriptsize int}}
&=&M^{2}(\Lambda )\phi^{\dagger}\phi 
+\frac{\lambda_E(\Lambda)}{6}(\phi^{\dagger}\phi )^2
+e^{2}_{E}(\Lambda)\phi^{\dagger}\phi A_{i}^{2}
+h_E^2(\Lambda)\phi^{\dagger}\phi\rho^{2}\\
&&+ie_{E}(\Lambda)A_{i}
(\phi^{\dagger}\partial_{i}\phi-\phi\partial_{i}
\phi^{\dagger})+\frac{\lambda_A(\Lambda)}{24}\rho^4+\delta{\cal L}.
\eqa
Using strict perturbation theory the Lagrangian of MSQED is split
in a way that is now familiar: 
\bqa \nonumber
({\cal L}_{\mbox{\scriptsize MSQED}})_{0}&=&
\frac{1}{4}\tilde{F}_{ij}\tilde{F}_{ij}+
(\partial_{i}\tilde{\phi} )^{\dagger}(\partial_{i}\tilde{\phi} )
+{\cal L}_{\mbox{\footnotesize gf}}+
{\cal L}_{\mbox{\footnotesize gh}},\\ 
\nonumber
({\cal L}_{\mbox{\scriptsize MSQED}})_{\mbox{\scriptsize int}}
&=&
\tilde{M}^{2}(\Lambda)
\tilde{\phi}^{\dagger}\tilde{\phi}+e_M^2(\Lambda)
\tilde{\phi}^{\dagger}\tilde{\phi}\,\tilde{A}_{i}^{3d}\tilde{A}_{i}^{3d}
+ie_M(\Lambda)\tilde{A}_{i}
(\tilde{\phi}^{\dagger}\partial_{i}\tilde{\phi}-
\tilde{\phi}\partial_{i}
\tilde{\phi}^{\dagger})\\
&&
+\frac{\lambda_{M}(\Lambda)}{6} 
(\tilde{\phi}^{\dagger}
\tilde{\phi})^{2}+\delta{\cal L}^{\prime}.
\eqa
In MSQED, wiggly and solid lines denote the propagators
of photons and charged scalars, respectively.
Again, we only show the diagrams which are nonzero in the Landau gauge.
\subsection{Field Normalization Constants}\label{felt2}
In the tree approximation the fields in ESQED and MSQED
are related as
\beq
\label{tree}
\tilde{\phi}(\Lambda)\approx\phi(\Lambda),\hspace{1cm}\tilde{A}_{i}(\Lambda)
\approx A_{i}(\Lambda).
\eeq
Again the field normalization constant can be read off from the momentum
dependent part of the propagator of
the underlying theory, which in this case is ESQED.
Consider first the scalar field. In strict perturbation
theory it is consistent to make a series expansion of the propagator
in powers of the external momentum ${\bf k}$. 
The only mass scale provided in the
loop integrals of ESQED is then the mass $m^2_E(\Lambda)$. Since the only
one--loop diagram involving the field $\rho$ 
is independent of the external momentum
(the tadpole in Fig.~\ref{hjort}), 
the one--loop correction to the 
momentum dependent part of the propagator vanishes. Hence there is no
renormalization of the 
field $\tilde{\phi}$ at one--loop.

A similar argument holds for the gauge field
and so~(\ref{tree}) holds to next--to--leading order.
\subsection{Coupling Constants} 
In this subsection we determine the gauge coupling $e^2_M(\Lambda)$
and the quartic coupling $\lambda_M(\Lambda)$ to next--to--leading order
in the parameters  of ESQED. The results for $N=1$ appear in Ref.~\cite{rummu}.
The matching is somewhat simplified, since the fields in MSQED can be directly
identified with the fields in ESQED to next--to--leading order.

Consider first the coupling constant $\lambda_M(\Lambda)$.
The quartic coupling $\lambda_M(\Lambda)$ is determined by the 
the following matching equation, in complete analogy with the 
calculations of $\lambda_E(\Lambda)$ in subsection~\ref{previ},
\label{ana}
\beq
\label{lmqed}
\Gamma_{\phi_1\phi_1\phi_1\phi_1}^{\mbox{\scriptsize ESQED}}(0)=
\Gamma_{\phi_1\phi_1\phi_1\phi_1}^{\mbox{\scriptsize MSQED}}(0).
\eeq
The only one--loop diagram contributing to 
$\Gamma_{\phi_1\phi_1\phi_1\phi_1}^{\mbox{\scriptsize ESQED}}({\bf k})$ 
at zero external momenta is
displayed in Fig.~\ref{tryte} and the correlator is:
\bqa
\Gamma_{\phi_1\phi_1\phi_1\phi_1}^{\mbox{\scriptsize ESQED}}(0)&=&
\lambda_E(\Lambda)-6e^{4}_{E}\int_{p}\frac{1}{(p^{2}+m_{E}^{2})^{2}}
\eqa
In MSQED, the one--loop corrections to the correlator 
$\Gamma_{\phi_1\phi_1\phi_1\phi_1}^{\mbox{\scriptsize MSQED}}({\bf k})$ 
vanish.
Using the matching equation (\ref{lmqed})
and appendix B, we finally end up 
with
\bqa
\lambda_M(\Lambda)=\lambda_E(\Lambda)-\frac{3e^{4}_{E}}{4\pi m_{E}}.
\eqa
Here, $\lambda_E(\Lambda)$ is given by~(\ref{lambdae}).

Consider next the gauge coupling $e^2_M(\Lambda)$. 
This coefficient is determined by 
calculating the 
correlator $\Gamma_{\Phi_1\Phi_1A_iA_j}^{\mbox{\scriptsize ESQED}}(0)$ 
and matching with the
corresponding correlator in MSQED,
$\Gamma_{\Phi_1\Phi_1A_iA_j}^{\mbox{\scriptsize MSQED}}(0)$.
It is easy to show that there is no one--loop correction to the result
from matching at tree--level, and so we have
\beq
e_{M}^{2}(\Lambda)=e_{E}^{2}(\Lambda),
\eeq
where $e^2_E(\Lambda)$is given by~(\ref{ekopling}).
\subsection{Mass Parameter}
In this subsection we determine the scalar mass parameter in the two--loop
approximation. For $N=1$ this has previously been carried out in~\cite{rummu}.
We calculate $\tilde{M}^2(\Lambda)$ by matching the
Higgs propagator in ESQED, 
$\Gamma_{\phi_1,\phi_1}^{\mbox{\scriptsize ESQED}}({\bf k})$    
with the Higgs propagator in MSQED,
$\Gamma_{\phi_1,\phi_1}^{\mbox{\scriptsize MSQED}}({\bf k})$.
The diagrams contributing to the 
two--point function in the strict perturbation expansion of ESQED 
are displayed in
Fig.~\ref{hjort}.
After Taylor expanding the self--energy function in ESQED in powers of 
$k^2$, the mass $m^2_E(\Lambda)$ is the only mass scale in the loop
diagrams. This implies that all loop diagrams which does not involve the
field $\rho$ vanish in dimensional regularization. The two--point function
in ESQED then reads
\bqa\nonumber
\label{emasse}
\Gamma_{\phi_1,\phi_1}^{\mbox{\scriptsize ESQED}}({\bf k})&=&
k^2+M^2(\Lambda)+
e^{2}_{E}\int_{p}\frac{1}{p^{2}+m_E^{2}}
\\&&
-2e_{E}^{4}\int_{pq}\frac{1}{(p^{2}+m_E^{2})(q^{2}+m_E^{2})({\bf p}+{\bf k}-{\bf q})^{2}}.
\eqa
The one--loop diagrams in MSQED
are the same as in ESQED, except for those diagrams which involve the
real scalar field $\rho$. Thus, the loop corrections to 
scalar self--energy function in strict perturbation theory, vanish
and 
the only non--vanishing contribution comes from the 
mass counterterm $\delta \tilde{M}^2$, 
which cancels the logarithmic ultraviolet divergences
\beq
\label{mmasse}
\Gamma_{\phi_1,\phi_1}^{\mbox{\scriptsize MSQED}}({\bf k})
=k^2+\tilde{M}^2(\Lambda)+\delta\tilde{M}^2.
\eeq
Matching the two expressions,~(\ref{emasse}) and~(\ref{mmasse})
we find
\bqa
\tilde{M}^2(\Lambda)&=&M^2(\Lambda)+
e^{2}_{E}\int_{p}\frac{1}{p^{2}+m_E^{2}}-
2e_{E}^{4}\int_{pq}\frac{1}{(p^{2}+m_E^{2})(q^{2}+m_E^{2})({\bf p}-{\bf q})^{2}}
-\delta\tilde{M}^2.
\eqa
The integrals are tabulated in Appendix B. The two--loop integral is 
ultraviolet divergent. The pole in $\epsilon$ must then be canceled
by the mass counterterm, which is
\beq
\delta \tilde{M}^{2}=\frac{e_{E}^{4}}{2(4\pi )^{2}\epsilon}.
\eeq
Our final expression for the scalar mass parameter in MSQED is
\bqa
\tilde{M}^{2}(\Lambda)&=&M^2(\Lambda)-\frac{e^{2}_{E}Tm_E}{4\pi}
-\frac{e_{E}^{4}}{16\pi^{2}}\left[1+2\ln\frac{\Lambda}{2m_E}\right].
\eqa
Here, the mass parameter $M^2(\Lambda)$ is given by~(\ref{emass}).
\section{Summary}
In the present paper we have discussed the dimensional
reduction approach to hot field theories which has been developed into
a quantitative tool by Farakos {\it et al.}~\cite{rummu} and by Braaten and 
Nieto~\cite{braaten}.
The basic idea is to exploit the fact there are 
two or more well--separated
mass scales in the system and that the heavy degrees of freedom 
decouple at long distance leaving us with effective field theories
of the light degrees of freedom. The effects of the heavy modes are
to renormalize the parameters in the effective theory and to induce new higher
order interactions.

In this work I have applied this method to a field theory consisting
of $N$ charged scalars coupled to an Abelian gauge field.
I have presented the
calculations of the parameters of ESQED and MSQED to next--to--leading
order in the parameters $\nu^2$, $\lambda$  and $e^2$ of full SQED.
The results are a generalization of 
existing results for $N=1$~[21,36].

The effective field theory (MSQED) that we have obtained can now be
used for a non--perturbative study of the phase transition on the lattice.
This includes in particular the order of the phase transition as a function
of the number of scalar fields $N$. 

\section*{Acknowledgments}
The author would like to thank Eric Braaten for useful discussions
and suggestions.
\appendix\bigskip\renewcommand{\theequation}{\thesection.\arabic{equation}}
\setcounter{equation}{0}\section{Sum--integrals in the Full Theory}
Throughout the work we use the imaginary 
time formalism, where the four--momentum is $P=(p_{0},{\bf p})$
with $P^{2}=p_{0}^{2}+{\bf p}^{2}$. 
The Euclidean energy takes on discrete values, $p_{0}=2n\pi T$
for bosons. 
Dimensional regularization is used to
regularize both infrared and ultraviolet divergences by working
in $d=4-2\epsilon$ dimensions, 
and we apply the $\overline{MS}$ 
renormalization scheme. We shall use the following notations for the 
sum--integrals that appear
\bqa
\hbox{$\sum$}\!\!\!\!\!\!\int_Pf(P)&\equiv &\left( \frac{e^{\gamma_{\tiny E}}\mu^{2}}
{4\pi}\right)^{\epsilon}\,\,\,\,
T\!\!\!\!\!
\sum_{p_{0}=2\pi nT}\int\frac{d^{3-2\epsilon}p}
{(2\pi)^{3-2\epsilon}}f(P).
\eqa
The one--loop sum--integrals needed in this work have been calculated in 
e.g. Ref.~\cite{arnold1}:
\bqa
\hbox{$\sum$}\!\!\!\!\!\!\int_P\frac{1}{P^{2}}&=&\frac{T^{2}}{12}
\left(\frac{\mu}{4\pi T}\right)^{2\epsilon}
\left[1+\left(2+2\frac{\zeta^{\prime}(-1)}{\zeta (-1)}\right)
\epsilon +{\cal O}(\epsilon^{2})\right],\\
\hbox{$\sum$}\!\!\!\!\!\!\int_P\frac{1}{(P^{2})^{2}}&=&\frac{1}{16\pi^{2}}
\left(\frac{\mu}{4\pi T}\right)^{2\epsilon}
\left[\frac{1}{\epsilon}+2\gamma_{E}+{\cal O}(\epsilon )\right],\\
\hbox{$\sum$}\!\!\!\!\!\!\int_P\frac{P_{0}^{2}}{(P^{2})^{2}}&=&-\frac{T^{2}}{24}
\left(\frac{\mu}{4\pi T}\right)^{2\epsilon}
\left[1+2\frac{\zeta^{\prime}(-1)}{\zeta (-1)}\epsilon
+{\cal O}(\epsilon^{2} )\right],\\
\hbox{$\sum$}\!\!\!\!\!\!\int_P\frac{P_{0}^{2}}{(P^{2})^{3}}&=&
\frac{1}{64\pi^{2}}\left(\frac{\mu}{4\pi T}\right)^{2\epsilon}\left[\frac{1}{\epsilon}+2+2\gamma_{E}+{\cal O}(\epsilon)\right],\\
\label{show}
\hbox{$\sum$}\!\!\!\!\!\!\int_P\frac{P_{0}^{4}}{(P^{2})^{4}}&=&
\frac{1}{128\pi^{2}}\left(\frac{\mu}{4\pi T}\right)^{2\epsilon}\left[\frac{1}{\epsilon}+\frac{8}{3}+2\gamma_{E}+{\cal O}(\epsilon)\right].
\eqa
Here, $\gamma_E$ is the Euler--Mascharoni constant and $\zeta (x)$ is the
Riemann Zeta function.\\ \\
The only two--loop graph needed
has been calculated in e.g.~\cite{arnold1}:
\bqa
\hbox{$\sum$}\!\!\!\!\!\!\int_{PP}\frac{1}{P^{2}Q^{2}(P+Q)^{2}}&=&0.
\eqa  
\setcounter{equation}{0}\section{Integrals in the three Dimensional Theory}
In the effective three--dimensional theory we use dimensional regularization
in $3-2\epsilon$ dimensions to regularize infrared and ultraviolet 
divergences.
In analogy with Appendix A, we define
\beq
\int_{p}f(p)\equiv\left( \frac{e^{\gamma_{\tiny E}}\mu^{2}}
{4\pi}\right)^{\epsilon}\int\frac{d^{3-2\epsilon}p}
{(2\pi)^{3-2\epsilon}}f(p).
\eeq
Again $\mu$ coincides with the renormalization scale in the 
modified minimal subtraction renormalization
scheme.\\ \\
In the effective theory we need the following one--loop integrals
\bqa
\label{effloop}
\int_{p} \frac{1}{p^{2}+m^{2}}&=&-\frac{m}{4\pi}
\left[1
+{\cal O}(\epsilon)\right],\\
\label{elgen}
\int_{p} \frac{1}{(p^{2}+m^{2})^{2}}&=&\frac{1}{8\pi m}
\left[1
+{\cal O}(\epsilon)\right].
\eqa
The specific two--loop integral needed is
\bqa \nonumber
\int_{pq} \frac{1}{(p^{2}+m^{2})(q^{2}+m^{2})({\bf p}-{\bf q})^{2}}
&=&\frac{1}{16\pi^2}
\left[\frac{1}{4\epsilon}
+\frac{1}{2}+\ln\frac{\mu}{2m}+{\cal O}(\epsilon)\right].
\eqa
The above integrals have been computed by several authors, e.g. in 
Refs. [6,22,25].
 
\end{document}